\documentclass[prb,aps,floats,amssymb,showkeys,showpacs,preprint,
titlepage,bibnotes,tightenlines]{revtex4}
\usepackage{graphicx}
\usepackage{graphics}
\usepackage{epsfig,bm}
\usepackage{rotating}
\begin{document}

\title{
Triviality problem and the high-temperature expansions of\\
 the higher susceptibilities for the Ising and  the scalar field\\
 models on four-, five- and six-dimensional lattices\\ }

\author{P. Butera}

\email{paolo.butera@mib.infn.it}

\affiliation
{Dipartimento di Fisica Universita' di Milano-Bicocca\\
and\\
Istituto Nazionale di Fisica Nucleare \\
Sezione di Milano-Bicocca\\
 3 Piazza della Scienza, 20126 Milano, Italy}
 
\author{M. Pernici} 

\email{mario.pernici@mi.infn.it}

\affiliation
{Istituto Nazionale di Fisica Nucleare \\
Sezione di Milano\\
 16 Via Celoria, 20133 Milano, Italy}

\date{\today}
\begin{abstract}
High-temperature expansions are presently the only viable approach to
the numerical calculation of the higher susceptibilities for the spin
and the scalar-field models on high-dimensional lattices.  The
critical amplitudes of these quantities enter into a sequence of
universal amplitude-ratios which determine the critical equation of
state.  We have obtained a substantial extension through order 24, of
the high-temperature  expansions of the free energy (in
presence of a magnetic field) for the Ising models with spin 
$s \geq 1/2$ and for the lattice scalar field theory with quartic
self-interaction, on the simple-cubic and the body-centered-cubic
lattices in four, five and six spatial dimensions.  A numerical
analysis of the higher susceptibilities obtained from these
expansions, yields results consistent with the widely accepted ideas,
based on the renormalization group and the constructive approach to
Euclidean quantum  field theory, concerning the no-interaction
(``triviality'') property of the continuum (scaling) limit of spin-$s$
Ising and lattice scalar-field models at and above the upper critical
dimensionality.
\end{abstract}
\pacs{ 03.70.+k, 05.50.+q, 64.60.De, 75.10.Hk, 64.70.F-, 64.10.+h}
\keywords{Scalar field triviality, Ising model, magnetic field }
\maketitle

\section{Introduction}
 The renormalization group theory (RG) theory\cite{wk,zinn,zinnb}
 predicts the value $d_c=4$ for the upper critical dimensionality of
 the $N$-component lattice scalar-field theory and of the short-range
 classical Heisenberg $N$-component spin systems with $O(N)$-symmetric
 interaction.  When $d \geq 4$, the critical fluctuations become too
 weak to drive the leading critical exponents away from the
 ``classical'' values taken in the mean field (MF) approximation, and
 can only induce minor corrections to scaling.  In particular, in 4D
 the simple MF asymptotic forms of the thermodynamical quantities at
 criticality should be corrected by logarithmic factors, whose precise
 structure is also predicted by the RG. In higher dimensions, the
 dominant singularities have  purely MF forms and the
 fluctuations can only influence the critical amplitudes and the
 corrections to scaling. These RG predictions entail the
 ``triviality''\cite{wk,sok,sokff,sym,calla} of the quantum
 $N$-component scalar-field theories in $d \geq d_c $, or, more
 precisely, the property that the continuum (scaling) limit of the lattice
 approximation of the theories (or of the spin models) describes
 fields whose connected fourth- and higher-order correlation-functions
 vanish and therefore are free or generalized-free.

  The main clues of this no-interaction property had been pointed out
 long ago\cite{land}, but more stringent arguments were produced only
 by the modern developments of the RG theory\cite{wk,zinn,zinnb}.  In
 the same years, a rigorous constructive
 approach\cite{sok,sokff,aize,fro,cara,hara,shlo} based on the
 representation\cite{sym} of the lattice scalar-field as a gas of
 polymers, made it possible to prove conclusively that the continuum
 Euclidean quantum field theory, built as the scaling limit of a
 lattice theory (with the simplest nearest-neighbor discretization of
 the Laplacian) in the symmetric phase, is
 ``non-trivial''\cite{eck,brydg} in $d\leq 3 $ and ``trivial'' in
 $d\geq 5 $ dimensions.

  The rigorous results that exist  in 4D (and, in general,
 for $N>4$) are still incomplete, 
 although they strongly suggest that nevertheless the
 triviality property still holds. Therefore some room is left not only
 to numerical studies, but also to a variety of
 efforts\cite{galla,cons,suslov} (and the related
 controversies\cite{balo}), aimed either to exploit possible gaps in
 the arguments, or to relax some of the hypotheses underlying the
 constructive approach, in order to make the definition of a
 ``non-trivial'' continuum theory feasible.

 For $d \geq 4$, the MonteCarlo (MC) simulation approach to the
 numerical verification of the RG predictions is not yet completely
 satisfactory. The detailed exploration of the near-critical behavior
 is hampered by the necessity of considering systems of very large
 sizes, and in particular, at $d=4$, by the difficulty of an accurate
 characterization of the slowly varying logarithmic deviations from MF
 behavior. For $d \geq 4$, also the finite-size-scaling theory and the
 confluent corrections to scaling have been debated\cite{dohm,
 stauf,luij,akte,merd,jones}.  Thus relatively few of the numerous
 available MC
 studies\cite{wein,fox,dru,bern,frick,kim,grass,kenna,cons,luij,bitt,lundow}
 are likely to be extensive enough to yield a satisfactory overall
 description of these systems at criticality, in spite of the
 remarkable progress in the simulation algorithms with reduced
 critical slowdown.

On the other hand, for these systems high-temperature (HT) expansions
 have been until now derived only for a small number of observables
 and are too short\cite{fishga,gaunt,bakerkinc}, or perhaps barely
 adequate\cite{luscher,vlad,bcklein,staufad,hell} to extract reliable
 information in the critical region.  We believe however, that the HT
 series methods might bring further insight into this context,
 provided that for a conveniently enlarged set of observables, the
 lengths of the expansions can be significantly extended.  Recently,
 new stochastic algorithms\cite{prokof,deng,berche,wolff} have shown
 promise of deriving extremely long, although approximate, HT
 expansions valid for finite-size lattice systems. The
 application\cite{wolff} of these methods also to the triviality
 problem is particularly interesting.
 
  The
traditional graphical
\cite{wortis,fishga,gaunt,bakerkinc,bcoenne,bcg4,bcisiesse,butper,
nickelrehr,bcfi4,luscher}
or iterative\cite{deneef,arisue,bcm,bciter} methods of calculation, although
severely limited by the fast increase of their combinatorial
complexity with the order of expansion, remain necessary to derive the
exact HT series coefficients, valid in the thermodynamical limit,
which are needed for a reliable use of the known analytic extrapolation
tools\cite{hunterbaker,fishauy,guttda}, such as Pad\'e
approximants(PA) or differential approximants (DA).  It is finally worth
adding that, in the case of high-dimensional models, these exact HT
expansions still seem to be the only practicable method to compute the
higher-order field-derivatives of the free energy at zero field, usually called
``higher susceptibilities''.

In this paper, we focus on the HT series approach to provide 
 further numerical evidence supporting the RG predictions in the case
 of the $N=1$ lattice scalar-field models and of the Ising spin-$s$
 systems. For this purpose,  we have computed and analyzed
 exact HT expansions of the higher susceptibilities, through order 24,
 to study their critical behavior and an important class of universal
 combinations of critical amplitudes (UCCAs), whose properties
 might also be of interest.  

The paper is organized as follows. In Section II we define the
 spin-$s$ Ising and the lattice scalar-field systems for which we have
 substantially extended the HT expansions of the specific free-energy
 in the presence of a uniform magnetic field. Then we make due
 reference to the few HT data already in the literature.  In Section
 III, we introduce the higher susceptibilities and indicate how their
 expected critical behavior varies with the lattice dimensionality.
 In Section IV, we review the definition of the dimensionless
 $2n$-points renormalized coupling constants in terms of the higher
 susceptibilities and indicate their role in the discussion of the RG
 predictions.  Then we introduce several classes of UCCAs related to
 the latter quantities.  The following Section V is devoted to a
 detailed numerical analysis of our HT expansions including
 discussions of numerical estimates of the critical temperatures,
 exponents and several UCCAs for the models under scrutiny.  The final
 Section contains our conclusions.

\section{ Ising-type models. Definitions and notation}
In what follows, we shall be concerned only with spin-$s$ Ising or
 one-component lattice scalar-fields, so that, unless explicitly needed,
 it will be convenient to drop 
 the dependence of the physical quantities on the number $N$ of
 components of the spin or of the field.

In a bounded region $\Lambda \subset {\cal Z}^d$ 
of the $d$-dimensional lattice $ {\cal Z}^d$, 
the spin-$s$ Ising model interacting with an external uniform 
magnetic field $H$ is  described by the 
Hamiltonian\cite{wortis,nickelrehr,bcisiesse,butper}
\begin{equation}
{\cal H}_{\Lambda}\{s\}=
-\frac{J} {s^2}  \sum_{<ij>} s_is_j-\frac{mH}{s}\sum_i s_i
\label{isingesse}
\end{equation}
where $s_i=-s,-s+1,...,s$ is the spin variable at the lattice site
 $\vec i$, $m$ is the magnetic moment of a spin, $J$ is the exchange
 coupling.  Within the region $\Lambda$, the first sum extends over
 all distinct nearest-neighbor pairs of sites, the second sum over all
 lattice sites.  Clearly, the conventional Ising model is obtained
 simply by setting $s=1/2$.

 The self-interacting   one-component   scalar-field lattice model in a
 magnetic field is described by the
 Hamiltonian\cite{bakerkinc,luscher,bcfi4,butper}
\begin{equation}
{\cal H}_{\Lambda}\{\phi\}= 
-\sum_{<ij>} \phi_i \phi_j+\sum_i (V(\phi_i) + H\phi_i).
\label{pdifi}
\end{equation}
Here $-\infty <\phi_i< +\infty$ is a continuous variable associated to
 the site $\vec i$ and $V(\phi_i)$ is an even polynomial in the
 variable $\phi_i$. In this study, for brevity 
 we have only discussed the particular 
 model in which $ V(\phi_i)= \phi^2_i + g(\phi^2_i-1)^2$, but
 considering interactions of a more general form requires only simple
 changes in the computation.

The Gibbs specific free energy 
 ${\cal F}(K,h)$ is  defined as usual by
\begin{equation}
 {\cal F}(K,h)= -\lim_{|\Lambda| \rightarrow \infty}
\frac {1} {|\Lambda| k_BT} {\rm ln} Z_{\Lambda} (K,h)
=-\lim_{|\Lambda| \rightarrow \infty} \frac {1} {|\Lambda| k_BT}
 {\rm ln} \sum_{conf}
 {\rm exp}[-{\cal H}_{\Lambda}/k_BT] 
\end{equation}
  Here $|\Lambda|$ is the volume of the
 region $\Lambda$,
 $K=J/k_BT$ (or $K=1/k_BT$ in the case of the scalar-field models),
 called inverse temperature for short,
is the HT expansion variable, with $k_B$ the Boltzmann constant and $T$
the temperature, while $h=mH/k_BT$ denotes the reduced magnetic field.

We have studied the models described by eqs.(\ref{isingesse}) and
 (\ref{pdifi}) on the hyper-simple-cubic (hsc) and the
 hyper-body-centered (hbcc) lattices.  Following
 Ref.[\onlinecite{nickelrehr}], for $d \geq 4$, the hbcc lattices are
 defined as those in which the first neighbors $q \hat j$ of the site
 $\hat i$ are such that $ \hat i- \hat j=(\pm 1,\pm 1,...,\pm 1)$.
 This choice has the technical advantage, decisive for the
 computations on high dimensional lattices, that the ``lattice
 free-embedding numbers'', that enter into the contribution of each
 graph to the HT expansion, factorize so that they can be expressed as
 powers of those referring to the 1D lattice.  As a consequence of
 this drastic simplification, the computing time of the expansions is
 independent of the lattice dimensionality, whereas, in the case of
 the hsc lattices, it grows exponentially with the dimensionality.  We
 should also notice that, for $d>2$, the coordination number $q=2^d$
 of the hbcc lattice is much larger than the coordination number
 $q=2d$ of the hsc lattice and therefore the hbcc lattice expansions
 share the advantage of being notably smoother and faster converging
 than the hsc ones.

The expansions presented here are based on a calculation of the HT and
 low-field expansion of the free energy of various models described by
 the Hamiltonians eqs.(\ref{isingesse}) and (\ref{pdifi}), in presence
 of an external uniform magnetic field, that we have extended through
 the order 24.  In the case of the conventional Ising model, i.e. the
 model with spin $s=1/2$, such an expansion, through order 17, was
 already in the literature\cite{katsura,mckenzie} in the case of the
 four-dimensional hsc lattice (h4sc).  Our expansion agrees only up to
 order 16 with these data and, as a consequence, with the series
 coefficients of the ordinary susceptibility $\chi_2(K)$ and of the
 fourth field-derivative of the free energy $\chi_4(K)$, obtained from
 them and analyzed in Ref.[\onlinecite{gaunt}], as well as in some
 successive studies.  For the conventional Ising model, in addition to
 the expansion in the case of the h4sc lattice, we have also computed
 the analogous expansion in the case of the hbcc lattice in 4D
 (h4bcc).  For both the h4sc and the h4bcc lattices, we have moreover
 computed HT and low-field expansions in the case of the Ising models
 with spin $s=1,3/2,...,3$ and in the case of the Euclidean
 one-component scalar-field lattice models with an even-polynomial
 self-interaction.  We have finally repeated the series derivation for
 the same set of models in 5D and 6D, but restricting ourselves to the
 five-dimensional hbcc (h5bcc) and the six-dimensional hbcc (h6bcc)
 lattices, for the reasons of computational simplification indicated
 above. All these expansions do not exist in the literature.

 Altogether, we have examined these Ising-type models
in 28 cases distinct by spatial dimensionality, value of the 
 spin and structure of the lattice or of the 
interaction.  In a given dimension, all these models are
expected to belong to the same critical universality class and
therefore to be characterized by the same set of critical exponents
and UCCAs.

Finally, let us also mention  that our HT expansions for the Ising models
 in a magnetic field can be readily transformed into low-temperature (LT)
 and high-field expansions, from which  the spontaneous magnetization and 
 the LT higher susceptibilities  can be derived. 

In our calculation of the HT expansions, we have employed the
  linked-cluster graphical method of Ref.[\onlinecite{wortis}].  We
  have used an algorithm of graph generation and series calculation
  already described in Ref.[\onlinecite{butper}]. The details of the
  computer implementation of this procedure, its validation, and its
  performance are discussed in the same paper, that was devoted to a
  study of the higher susceptibilities and the scaling equation of
  state for the 3D Ising universality class.  Our extensions of the HT
  and low field expansions are summarized in Table \ref{tab1}. The set
  of series coefficients cannot fit into this paper because of its
  large size and will be tabulated elsewhere.

\subsection{Available  series expansions in zero field}

 It is appropriate to list here the few HT expansions of the higher
 susceptibilities for high-dimensional models at zero field that can
 already be found in the literature. All of them are restricted to the
 conventional spin-1/2 Ising model on the hsc lattices in zero field.
 The ordinary susceptibility $\chi_2(K)$ was derived\cite{fishga}
 through order 11 in dimensions $d=2,..,6$. More recently, these
 calculations were extended\cite{bakerkinc} to include, through the
 the same order, also $\chi_4(K)$ and the second moment of the
 correlation function $\mu_2(K)$, in $d=2,3,4$ dimensions and
 carried\cite{luscher} up to order 14. The expansion of the
 susceptibility $\chi_2(K)$ has been recently pushed\cite{hell} to
 order 19 on the h4sc and h5sc lattices. An expansion of $\chi_2(K)$
 valid for any dimension $d$ was computed\cite{gofman} through order
 15. For $\chi_2(K)$, $\chi_4(K)$ and the sixth field-derivative of
 the free energy $\chi_6(K)$, strong coupling expansions through order
 11, i.e. expansions in powers of the second-moment correlation length
 $\xi^2(K)=\mu_2(K)/2d\chi_2(K)$, instead of $K$, valid for any $d$,
 have also been obtained\cite{benderboett}. Of course, the usual HT
 expansions in powers of $K$ can be recovered simply by reverting the
 appropriate expansion of $\xi^2(K)$.

\begin{table}[ht]\scriptsize
\caption{ Maximal order in $K$ of the HT and low-field
 expansions of the free energy for the Ising and scalar-field
  models on four-, five- and six-dimensional lattices.  }
\begin{tabular}{|c|c|c|}
 \hline
           & Existing data\cite{mckenzie}   & This work\\
 \hline
${\rm h4sc}$ Ising $s=1/2$&        17 &     24\\
${\rm h4sc}$ Ising $s>1/2$&       0  &     24\\
${\rm h4sc}$ scalar field & 0 &     24\\
 \hline 
${\rm h4bcc}$ Ising $s \geq 1/2$&       0  &     24\\
${\rm h4bcc}$ scalar field & 0 &     24\\
 \hline 
${\rm h5bcc}$ Ising $s \geq 1/2$&       0  &     24\\
${\rm h5bcc}$ scalar field & 0 &     24\\
 \hline 
${\rm h6bcc}$ Ising $s \geq 1/2$&       0  &     24\\
${\rm h6bcc}$ scalar field & 0 &     24\\
 \hline
\colrule  
\end{tabular} 
\label{tab1}
\end{table}

\section{The HT expansions of the higher susceptibilities}
The assumption of asymptotic
scaling\cite{widom,dombhunter,pata,kada,fish67} for the singular part
${\cal F}_s (\tau,h)$ of the reduced specific free energy, valid 
for dimension $d \neq d_c$, when both $h$
and $\tau$ approach zero, is usually expressed in the form
\begin{equation}
{\cal F}_s (\tau,h) \approx |\tau|^{2-\alpha} Y_{\pm}(h/|\tau|^{\beta \delta}).
\label{freescaling}
\end{equation}

   where $\tau=(1-K/K_c)$ is the reduced inverse temperature.  The
   functions $Y_{\pm}(w)$ are defined for $0\le w \le \infty$ and the
   $+$ and $-$ subscripts indicate that different functional forms are
   expected to occur for $\tau<0$ and $\tau>0$. The exponent $\alpha$
   specifies the divergence of the specific heat, $\beta$ describes
   the small $\tau$ asymptotic form of the spontaneous specific
   magnetization $M$ on the phase boundary $(h \rightarrow 0^+,
   \tau<0)$
\begin{equation}
M \approx B(-\tau)^{\beta}
\label{ampmagsp}
\end{equation}
 and  $B$ denotes the critical amplitude of $M$. 
 The exponent 
 $\delta$ characterizes the small $h$ asymptotic behavior of the
magnetization on the critical isotherm $(h \ne 0,\tau=0)$, 
\begin{equation}
|M| \approx B_c |h|^{1/\delta}
\label{amphiso}
\end{equation}
 and $B_c$ is the corresponding critical amplitude.  For $d \geq d_c$,
the  MF values expected for the exponents $\alpha$, $\beta$ and $\delta$
are $\alpha=\alpha_{MF}=0$, $\beta=\beta_{MF}=1/2$ 
and $\delta=\delta_{MF}=3$, while for the
susceptibility exponent we have $\gamma=\gamma_{MF}=1$ and for the
correlation-length exponent $\nu=\nu_{MF}=1/2$.
  The usual scaling laws (but, of course, not the hyperscaling laws)
follow from eq.(\ref{freescaling}).

From our calculation of the magnetic-field-dependent free energy, we
have gained extensions of the  existing HT expansions in zero field and, in
addition, made available a large body of data not yet existing in the
literature, in particular for the $n$-spin connected correlation
functions at zero wave number and zero field
 (the ``higher susceptibilities''), defined by the
successive field-derivatives of the specific free energy
\begin{equation}
 \chi_{n}(K)=(\partial^n {\cal F}(K,h)/\partial h^n)_{h=0}
=\sum_{s_2,s_3,...,s_{n}}<s_1 s_2...s_{n}>_c.
\label{ncorr}
\end{equation}
 in the Ising model case,  
 or by the analogous formula in the scalar field case.  For odd values of
$n$, the quantities $\chi_{n}(K) $ 
 vanish in the symmetric HT phase, while they are
nonvanishing for all $n$ in the broken-symmetry LT
phase.

For even values of $n$ in the symmetric phase, the
RG theory predicts that, for $d>4$,  we have 
\begin{equation}
 \chi_{n}(\tau) \approx C^{+}_{n}|\tau|^{-\gamma_{n}}(1 +
 b^{+}_{n} |\tau|^{\theta} + \ldots).
\label{2ncorras5}
\end{equation}
as $\tau \rightarrow 0^+$ along the critical isochore ($h=0, \tau>0$).
In eq.(\ref{2ncorras5}), $b^{+}_n$ and $\theta$ denote, respectively,
 the amplitude and the exponent of the leading confluent correction to
 the asymptotic behavior. The
 explicit expressions obtained in the case of the spherical 
model\cite{joyce,figut,gutt} suggest that 
in 5D  one should expect $\theta=1/2$,
 whereas, in 6D, $\theta=1$, with possible multiplicative logarithmic
 correction terms.

At the marginal dimension $d_c$, the homogeneity property 
 described by eq.(\ref{freescaling}) is not
 strictly true, because of the expected logarithmic corrections.
In this case, for even values of $n$, in the symmetric phase, the
RG theory predicts for the higher susceptibilities the following 
 asymptotic behavior  
\begin{equation}
 \chi_{n}(\tau) \approx C^{+}_{n}\tau^{-\gamma_n} 
{\vert {\rm ln} (\tau)\vert^{G_n(N)}} 
\Big[1 + O\Big( {\rm ln}(\vert {\rm ln}(\tau)\vert)/{\rm ln}(\tau)\Big) \Big]
\label{2ncorras4}
\end{equation}
in the  $\tau \rightarrow 0^+$ limit.
In both eqs.  (\ref{2ncorras5}) and (\ref{2ncorras4}), one has
 $\gamma_{n}=\gamma_{MF} +(n-2)\Delta_{MF}$, with  the gap exponent
 $ \Delta_{MF}=\beta_{MF}\delta_{MF}=3/2$.  
 The general expression for $G_n(N) $ is 
\begin{equation}
G_n(N)= (\frac{3} {2} n-2)\frac{N+2} {N+8} -n/2 +1 
\label{gn}
\end{equation}
so that in the $N=1$ case, $G_n(1)=G=1/3$,
 independently of $n$.

Clearly, the usual hyperscaling relation 
$2\Delta=d\nu+\gamma$, which is valid for $d<d_c$, 
 fails by a power when $d>4$, while
 it is only logarithmically violated in $d=4$.

The simplest consequence of the usual scaling hypothesis
eq.(\ref{freescaling}), which will be tested using our HT expansions,
is that the critical exponents of the successive derivatives of ${\cal
F} (\tau,h)$ with respect to $h$ at zero field, are evenly spaced by
the gap exponent $\Delta_{MF}$.  Also in 4D, this property can be
simply and accurately checked by a HT analysis of the higher
susceptibilities.
\section{Renormalized couplings and related quantities}
 It is useful here to recall the definitions of the universal
quantities $g^+_{2n} $, called zero-momentum $n-$spin dimensionless
renormalized coupling constants (RCC's) in the symmetric phase. They
enter into the approximate representations of the scaling equation of
state and moreover play a key role in the discussion of the triviality
properties of the $d \geq 4$ systems. They are defined as the critical
limit when $K \rightarrow K_c^-$ of the expressions

\begin{equation}
 g_4(K)=-\frac{{\it v} } {\xi^d(K) } \frac{\chi_4(K)}
 { \chi_2^2(K)}
\label{g4}
\end{equation}
\begin{equation}
 g_6(K)=\frac{{\it v} ^2} {\xi^{2d}(K)}\Big [ -\frac{\chi_6(K)} {\chi_2^3(K)}
 + 10 \Big (\frac{\chi_4(K)} { \chi_2^2(K)}\Big)^2 \Big]
\label{g6}
\end{equation}
\begin{equation}
 g_8(K)=\frac{{\it v} ^3} {\xi^{3d}(K)}\Big [ -\frac{\chi_8(K)} 
{\chi_2^4(K)}
+56 \frac{\chi_6(K) \chi_4(K)} {\chi_2^5(K)}
 -280 \Big (\frac{\chi_4(K)} {\chi_2^2(K)}\Big)^3 \Big]
\label{g8}
\end{equation}

\begin{eqnarray}
 g_{10}(K)=\frac{{\it v} ^4} {\xi^{4d}(K)}
\Big [-\frac{\chi_{10}(K)} {\chi_2^5(K)}+
 120\frac{\chi_8(K) \chi_4(K)} {\chi_2^6(K)}
+126 \frac{\chi_6^2(K)} { \chi_2^6(K)}\\
\nonumber
 -4620 \frac{\chi_6(K) \chi_4^2(K)} { \chi_2^7(K)}
 +15400 \Big (\frac{\chi_4(K)} {\chi_2^2(K)}\Big)^4 \Big] 
\label{g10}
\end{eqnarray}

\begin{eqnarray}
  g_{12}(K)=\frac{{\it v} ^5} {\xi^{5d}(K)}
\Big [-\frac{\chi_{12}(K)} {\chi_2^6(K)}+
 220\frac{\chi_{10}(K) \chi_4(K)} {\chi_2^7(K)}
+792 \frac{\chi_8(K)\chi_6(K)} { \chi_2^7(K)}\cr
 -17160 \frac{\chi_8(K)\chi_4^2(K)} { \chi_2^8(K)}
-36036 \frac{\chi_6^2(K) \chi_4(K)} { \chi_2^8(K)}
 +560560 \frac{\chi_6(K)\chi_4^3(K)} {\chi_2^9(K)}\cr
-1401400\Big (\frac{\chi_4(K)} {\chi_2^2(K)}\Big)^5 \Big]
\label{g121}
\end{eqnarray}

and so on. The constant ${\it v} $ is a lattice-dependent geometrical
 factor called the volume per lattice site\cite{bcg2n}.  A longer list
 of the RCC's appears in Ref.[\onlinecite{butper}], where the equation
 of state is discussed only for the 3D case. For technical reasons, we
 have not yet extended the HT expansions of $\mu_2(K)$ and,
 correspondingly, of the second-moment correlation length $\xi(K)$, so
 that in this paper we shall study only the ratios of RCC's, for
 $n>2$,
\begin{equation}
r_{2n}(K)=\frac {g_{2n}(K)}{g_{4}(K)^{n-1}}
\label{r2n}
\end{equation}
  which share the computational advantage of being independent of
$\xi(K)$. The critical limits of these ratios are universal quantities 
that will be denoted by $r^+_{2n}$.  

At the upper critical dimension $d_c$, the quantities $ g_{2n}(K)$ are
expected to vanish like powers of $1/{\rm ln}(\tau)$, when $\tau
\rightarrow 0^+$. Therefore the continuum limit theory is consistent
only for vanishing renormalized coupling, i.e. it is trivial.  We can
check numerically that, in the same limit, the lowest ratios
$r_{2n}(K)$ remain finite in 4D.  For $d \geq 5$, both the $
g_{2n}(K)$ and the $r_{2n}(K)$ vanish in the critical limit like
powers of $\tau$, so that the mentioned property of triviality is also
true for $d>d_c$.

 Briefly recalling  more detailed discussions\cite{zinnb,guidazinn,butper}, 
we can also observe that, for
$d>4$, in the small magnetization region, where the reduced magnetic
field $ h(M,\tau) $ has a convergent expansion in odd powers of the
magnetization $M$, the critical equation of state can be written in
terms of an appropriate variable $z \propto M\tau^{-\beta}$ as
\begin{equation}
 h(M,\tau) = \bar h|\tau|^{\beta\delta}F(z)
\label{eqstat}
\end{equation}
where $\bar h$ is a constant and $F(z)$ is normalized by the equation 
$F'(0)=1$.
In general, the small $z$ expansion of $F(z)$ can be written as

\begin{equation}
 F(z)=z+\frac{1}{3!}z^3+\frac{r^+_6} {5!} z^5+\frac{r^+_8} {7!} z^7+...
\label{fz}
\end{equation}
  In the MF approximation,  all $r^+_{2n}$ vanish, and $F(z)$
 reduces to $F_{MF}(z)=z+\frac{1}{3!}z^3$.

At the upper critical dimension, the following form of the critical universal
 equation of state for an $N$-component system is obtained\cite{zinn,zinnb} 
 from the RG
\begin{equation}
H \propto \left [ M \tau |{\rm ln} M|^{(N+2)/(N+8)} 
+ \frac{1} {(N+8)} \frac{M^3}
{|{\rm ln}M |} \right ](1+\frac{const.} {|{\rm ln}M |})
\label{es4d}
\end{equation}
 valid for small $M$ and $H$.
From eq.(\ref{es4d}) the general formula eq.(\ref{2ncorras4}) and 
the expression (\ref{gn}) for $G_n(N)$  can be
 deduced.

In terms of the higher susceptibilities, the simple sequence of quantities
 was defined\cite{watson} long ago
 \begin{equation}
 {\cal I}_{2r+4}(K)=\frac{ \chi^r_2(K) \chi_{2r+4}(K)}{\chi^{r+1}_4(K)} 
\label{watso}
\end{equation}
 with $r \geq 1 $. The finite and universal critical values 
 \begin{equation}
{\cal I}^+_{2r+4}=\frac{(C_2^+)^r C_{2r+4}^+} {(C_4^+)^{r+1}}
\label{watsoncr}
\end{equation}
  of the functions ${\cal I}_{2r+4}(K) $ in the limit $K \rightarrow K_c^-$,
 include some of the UCCAs first described in the literature.  

Together
 with the sequence ${\cal I}^+_{2r+4}$ of UCCAs,  the sequences
 ${\cal A}^+_{2r+4}$ and ${\cal B}^+_{2r+8}$, obtained as the critical
 limits of the functions 
\begin{equation}
{\cal A}_{2r+4}(K)=\frac{\chi_{2r}(K)  \chi_{2r+4}(K) } {(\chi_{2r+2}(K) )^2}
\label{watsonar}
\end{equation}

\begin{equation}
{\cal B}_{2r+8}(K)=\frac{\chi_{2r}(K)  \chi_{2r+8}(K) } {(\chi_{2r+4}(K) )^2}
\label{watsonbr}
\end{equation}
  with $r \geq 1 $, 
 were  also defined in Ref.[\onlinecite{watson}].

In 4D the general formula eq.(\ref{gn}) for $G_n(N)$ implies that the
 powers of the logarithms that enter into the leading critical
 singularities eq.(\ref{2ncorras4}) cancel in the quantities 
 ${\cal I}_{2r+4}(K)$, ${\cal A}_{2r+4}(K)$, and ${\cal B}_{2r+8}(K)$ 
at the critical limit. Conversely, eq.(\ref{gn}) can also be obtained
 recursively from the knowledge of only $G_2(N)$ and $G_4(N)$ by
 requesting that such a cancellation occurs.
 
 The ratios  $r_{2n}(K)$  can be  simply expressed in terms of the 
functions  ${\cal I}_{2r+4}(K)$. For example
\begin{equation}
 r_6(K)=\frac {g_{6}(K)}{g_{4}(K)^{2}} = -{\cal I}_6(K)+10 
\label{r6}
\end{equation}
\begin{equation}
 r_8(K)=\frac {g_{8}(K)}{g_{4}(K)^{3}}= {\cal I}_8(K)-56 {\cal I}_6(K)+280
\label{r8}
\end{equation}
and so on. Taking the $K \rightarrow K_c^-$ limit in the
eqs.(\ref{r6}), (\ref{r8}), etc.  and observing that the quantities
$r_{2n}(K)$ vanish as $K \rightarrow K_c^-$ in the MF
approximation\cite{zinn}, the corresponding critical values 
$\hat{\cal I}^+_{2r+4}$ of the quantities in eq.(\ref{watso}) can be simply
evaluated, obtaining $\hat {\cal I}^+_6=10$, $\hat {\cal I}^+_8=280 $,
$\hat {\cal I}^+_{10}=15400 $, $\hat {\cal I}^+_{12}=1401400 $, etc.
It is also not difficult\cite{joyce} to compute the MF values of the
 first few terms of the 
sequences ${\cal A}^+_{2r+4}$ and ${\cal B}^+_{2r+8}$. For example:
$\hat {\cal A}^+_{8}=14/5$, $\hat {\cal A}^+_{10}=55/28$
$\hat {\cal B}^+_{10}=154$, and $\hat {\cal B}^+_{12}=143/8$.

In the next Section, we shall study numerically the first few terms of
the sequences ${\cal I}^+_{2r+4} $, ${\cal A}^+_{2r+4}$ and ${\cal
B}^+_{2r+8}$ and observe that they share similar properties.

\section{Results of the series analysis}
We address the reader to Refs.[\onlinecite{bcisiesse,butper}], for a
more detailed description of the numerical approximation techniques
necessary to estimate the critical parameters in the models under
study, i.e. the locations of the critical points, their exponents of
divergence and the critical amplitudes for the various
susceptibilities.  We shall employ the DA method, a
generalization\cite{hunterbaker,fishauy,guttda} of the elementary PA
method to resum the HT expansions nearby the border of their
convergence disks.  This technique consists in estimating the values
of the finite quantities or the parameters of the singularities of the
expansions of the singular quantities from the solution, called {\it
differential approximant}, of an initial value problem for an ordinary
linear inhomogeneous differential equation of the first- or of a
higher-order. The equation has polynomial coefficients defined in such
a way that the series expansion of its solution equals, up to an
appropriate order, the series under study.  In addition to this
technique, we shall also use a smooth and faster converging
modification\cite{zinnmra,guttda} of the traditional method of
extrapolation of the coefficient-ratio sequence, sometimes called
modified-ratio approximant (MRA) method, to determine the location and
the exponents of critical points.

Using PA or DA approximants, one can achieve, in some cases,
evaluations of the parameters which are unbiased, i.e.  obtained
without using independent estimates of the critical temperature in the
construction of the approximants.  In some cases however, accurate
estimates of the critical inverse temperatures are necessary to bias
the determination of the critical exponents and amplitudes.  As a
general comment on the uncertainties of the estimates obtained by
these methods, we have to observe that, inevitably, they are rather
subjective. Therefore, we should be very cautious, compare the
estimates obtained from the different approximation methods and also
check how effectively our tools perform when applied to artificial
model functions having the expected singularity structure. In our DA
calculations, the uncertainties are taken as a small multiple of the
spread among the estimates obtained from an appropriate sample of the
highest-order approximants i.e. those using most or all available
expansion coefficients. Similarly, in the case of the MRAs, the error
bars will be defined as a small multiple of the uncertainty of an
appropriate extrapolation\cite{bcisiesse} of the highest-order
approximants.
\begin{table}
\caption{ Our estimates of the critical inverse temperatures of the 
 spin-$s$ Ising models for various lattices and several values of the spin 
in 4D, in 5D and in 6D. }
\begin{tabular}{|c|c|c|c|c|c|c|}
 \hline
  lattice  &$s=1/2$  &$s=1 $ &$s=3/2$ & $s=2$&$s=5/2$ &$s=3$  \\
 \hline 
${\rm h4sc}$ &0.149693(3)&0.215597(3)&0.255641(3)&0.282568(3)&0.301919(3)
&0.316497(3)\\
 \hline
${\rm h4bcc}$&0.0690114(8)&0.101165(2)&0.120592(2)&0.133605(2)&0.142930(2)
&0.149942(2)\\
 \hline
${\rm h5bcc}$&0.0326478(3) &0.0484554(3)&0.0579714(3)&0.0643290(3)&0.0688769(3)
 &0.0722915(3) \\
 \hline
${\rm h6bcc}$&0.0159390(2) &0.0237914(2)  &0.0285102(2) 
 &0.0316592(2)  &0.0339099(2)  &0.0355989(2)  \\
 \hline
\colrule  
\end{tabular} 
\label{tabkc}
\end{table}
\begin{table}
\caption{ Our estimates of the critical inverse temperatures of the 
 scalar-field  models for various lattices and several values
 of the quartic self-coupling in 4D, in 5D and in 6D.}
\begin{tabular}{|c|c|c|c|c|c|c|}
 \hline
  lattice    &$g=0.5 $ &$g=0.6$ & $g=0.9$&$g=1.1$ &$g=1.3$&$g=1.5$  \\
 \hline 
${\rm h4sc}$& 0.283025(3)&0.280704(3) &0.270597(3)  &0.262806(3)&0.254915(3)
 &0.247233(3) \\
 \hline
${\rm h4bcc}$&0.136451(2)&0.134940(2)&0.129177(2) &0.124991(2)
&0.120852(2) &0.116885(2)\\
 \hline
${\rm h5bcc}$&0.0665777(2)&0.0657078(2)&0.0625961(2)&0.0604114(2)
&0.0582806(2) &0.0562586(2)\\
 \hline
${\rm h6bcc}$&0.0329566(1)&0.0324976(1)&0.0308921(1)&0.0297796(1) 
&0.0287008(1)  &0.0276811(1)\\
 \hline
\colrule  
\end{tabular} 
\label{tabkcfi4}
\end{table}
\subsection{The critical inverse temperatures of the models}
If in 4D, as predicted by the RG theory, logarithmic factors modify
the structure of the leading critical singularities and also appear in
the corrections to scaling,  as described by eq.(\ref{2ncorras4}), we
should expect that the numerical procedures mentioned above might
suffer from a slower convergence than in the case of pure power-law
scaling.  For the determination of the critical
temperatures, different approximation methods such as PAs, DAs and
extrapolated MRAs have been used to study the expansions of the ordinary
susceptibility $\chi_2(K)$, the quantity which generally shows the
fastest convergence. 
Independently of the lattice type and dimensionality, our best
estimates of the critical inverse temperatures for the systems under
study are obtained extrapolating to large order $r$ of expansion, 
 a few (from four to seven)
highest order terms of the MRA sequences of estimates $(K_c)_r$ of the
 critical inverse temperatures. 
To perform the extrapolation,
we rely on the validity\cite{bcisiesse} of the simple asymptotic form
\begin{equation}
(K_c)_r=K_c- \frac{\Gamma(\gamma)} {\Gamma(\gamma-\theta)} 
\frac{\theta^2(1-\theta)b_2}{r^{1+\theta}}+o(\frac{1}{r^{1+\theta}}).
\label{kras}
\end{equation}
In general, $\theta$ and $b_2$ indicate respectively the exponent and
the amplitude of the leading confluent correction to scaling of
$\chi_2(K)$.

In the 4D case, in which the asymptotic critical behavior of
 $\chi_2(K)$ is described by eq.(\ref{2ncorras4}), we can take
 $\theta=0$. Therefore the second term on the right-hand side of
 eq.(\ref{kras}) vanishes and it must be replaced by a higher-order
 term depending on the exponent of the next-to-leading correction to
 scaling in eq.(\ref{2ncorras4}).  A similar argument applies in the 6D
 case in which we expect $\theta=1$.  In the 5D case, in which we
 expect $\theta=1/2$, the coefficient of $1/r^{1+\theta}$ in
 eq.(\ref{kras}) also appears to be negligible, so that the situation
 is similar to that of the 4D and the 6D cases. Since we do not know
 the values of the exponents of the next-to-leading correction to
 scaling, the simplest procedure of extrapolation might consist in
 assuming an asymptotic form $(K_c)_r=K_c+ w/r^{1+\epsilon}$ and in
 determining $K_c$, $w$, and $\epsilon$ by a best fit to our data.  We
 obtain the values $\epsilon= 0.6(2)$ in 4D, $\epsilon= 1.1(2)$ in 5D
 and $\epsilon= 1.5(2)$ in 6D. These estimates are compatible with our
 previous remarks, indicating that the asymptotic behavior of
 eq.(\ref{kras}) is determined by the next-to-leading rather than the
 leading correction to scaling.  At the same time, as suggested by
 M.E. Fisher\cite{fpc}, the expectations\cite{joyce,figut,gutt}
 concerning the exponent of the leading corrections to scaling, whose
 amplitudes are probably not negligible in spite of the fact that they
 are not seen by the MRAs, can be essentially confirmed studying by
 DAs the critical behavior of quantities like ${\cal I}_{2r+4}(K)$,
 ${\cal A}_{2r+4}(K)$, ${\cal B}_{2r+8}(K)$ etc. and of their
 derivatives. As above remarked, in these quantities the dominant
 critical singularities cancel, while the leading corrections to
 scaling should survive and could be detected by DAs.  In particular,
 a study of the derivatives of ${\cal I}_{6}(K)$ and ${\cal
 I}_{8}(K)$, for the spin-$s$ Ising models, leads to the values
 $\theta=0.25(10)$ in 4D, $\theta=0.45(10)$ in 5D and
 $\theta=0.95(10)$, in very reasonable agreement with the
 predictions\cite{joyce,figut,gutt}.

 Our final results for the critical inverse temperatures of some
 spin-$s$ Ising and scalar-field models are collected in the Tables
 \ref{tabkc} and \ref{tabkcfi4}.  In the 4D case, we have attached
 particularly generous error bars to our estimates.  In $d>4$
 dimensions, no logarithmic factors are expected to modify the leading
 MF behavior of the physical quantities, so that our approximation
 tools are likely to yield estimates of a higher accuracy, which
 moreover appear to improve with increasing lattice dimensionality,
 both because of the decreasing size of the corrections to scaling and
 of the increasing lattice coordination number. All these results are
 confirmed also by the analyses employing DAs.

Only in the case of the Ising model with spin $s=1/2$ on the h4sc
lattice,  we can compare our estimates with those obtained in other 
studies by extrapolation of shorter HT series.  In
Ref.[\onlinecite{staufad}] the estimate $K_c=0.149696(4)$ was obtained
from a series of order 17, while in Ref.[\onlinecite{hell}] the result
$K_c=0.149691(3)$ was derived from a series of order 19. As far as the
most recent large-scale MC simulations are concerned, the estimate
$K_c=0.149697(2)$ was obtained in Ref.[\onlinecite{staufad}], the
value $K_c=0.149697(2) $ in Ref.[\onlinecite{bitt}], while the value
$K_c=0.1496947(5) $ is reported in Ref.[\onlinecite{lundow}].  Our
result in Table \ref{tabkc} is fully consistent with the older estimates.
  No comparison is
possible either for higher values of the spin on the h4sc lattice, or
for any value of the spin on the h4bcc lattice, since no studies are
available for these systems.  
In the case of the higher dimensional lattices our
analysis includes only the h5bcc and h6bcc lattices, which have not
been studied elsewhere until now.

\subsection{ The logarithmic corrections in 4D}

Also in the computation of the critical exponents, it is convenient to
distinguish the 4D case from the higher dimensional ones.  

In 4D, when computing the exponent $\gamma$ of $\chi_2(K)$ by PAs or
 DAs, we obtain estimates very near to, but slightly larger than
 unity.  These estimates should then be regarded as the values of 
 ``effective exponents'' which reflect the presence of a small
 correction to the leading classical behavior (and of subleading
 corrections).  If we assume that the leading correction to MF
 behavior has the multiplicative logarithmic structure predicted by
 the RG, we can resort to a variety of procedures
 proposed\cite{guttda,gutt,adler,vlad} in the literature to isolate
 the logarithmic factor from the main power behavior and to measure
 its exponent.  These prescriptions generally amount to cancel out the
 main power-singularity in favor of the weak logarithmic one and
 therefore they need to be biased with an estimate of the inverse
 critical temperature, to which, in turn, the values obtained for the
 exponent of the logarithm are very sensitive.
 
\begin{figure}[tbp]
\begin{center}
\leavevmode
\includegraphics[width=3.37 in]{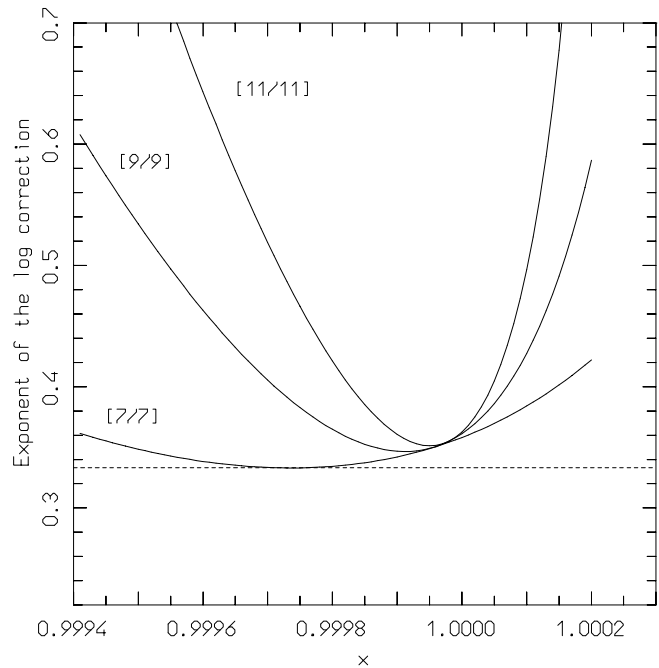}
\caption{ \label{figesplog} Various PA approximants
 ( [7/7], [9/9] and [11,11])  of the auxiliary function 
 $l(K;\tilde K_c)$ vs the bias parameter $x=\tilde K_c/0.101165$ 
normalized to our MRA estimate of $K_c$ in case of the 
spin $s=1$ model on the h4bcc lattice. 
The dashed horizontal line represents the predicted value of the 
exponent of the logarithmic correction to the leading power behavior of
the susceptibility.}
\end{center}
\end{figure}

 For example, in the case of the ordinary susceptibility $\chi_2(K)$,
  one might study the auxiliary function $l(K;\tilde K_c)$ defined by
\begin{equation}
l(K;\tilde K_c)=-(\tilde K_c-K){\rm ln}
(\tilde K_c-K)\frac{d}{dK}{\rm ln}[(\tilde K_c-K)\chi_2(K)]
\label{glog}
\end{equation}
 where $\tilde K_c$ is some accurate approximation of the true $K_c$.
By eq.(\ref{2ncorras4}), $l(K; K_c) = G +O({\rm ln|ln}\tau|)$, i.e. it
yields the value of the exponent $G$ when $K \rightarrow K_c^-$ and
$\tilde K_c= K_c$.  Since $\tilde K_c$ enters as a parameter into the
definition of this biased indicator, we should consider how the
estimate $G(\tilde K_c)$ of the exponent depends on the choice of
$\tilde K_c$ in a small vicinity of our best estimate of the critical
inverse temperature reported in Tabs.\ref{tabkc} or \ref{tabkcfi4}.
As a typical example, we show in Fig.\ref{figesplog} the plots of
$G(\tilde K_c)$ vs $\tilde K_c$ (normalized to our MRA estimate of $
K_c$), computed by PAs of various orders, in the case of the Ising
model with spin $s=1$ on the h4bcc lattice. It is reasonable to expect
that the value of $G(\tilde K_c)$ should depend slowly on $\tilde K_c$
near the exact value of the critical inverse temperature so that its
best value might perhaps correspond to a stationary point. We observe
that, for most PAs of $G(\tilde K_c)$, such a point does indeed exist
and also that the curves obtained by various PAs touch nearby this
point, which is generally not much different from our best estimate of
$K_c$ as reported in our Tables. In the
literature\cite{guttda,gutt,adler,vlad}, the value of $G(\tilde K_c)$
at the point where the various curves touch, is generally taken as the
most accurate estimate of the exponent $G$.  However, this choice  may
be questioned, since the result appears to be insensitive to the order of
approximation. As shown in Fig.\ref{figesplog}, if we take the value
of $G(\tilde K_c)$ at the stationary point as the best approximation,
the estimates are also close to the expected value $G=1/3$. 
 Unfortunately, also the choice of the stationary value as the best
approximation is open to doubt, since in this case the successive
approximations seem to worsen as the order of the series increases.
We must moreover mention that, in the h4bcc Ising system, the values
of $G$ computed in this way, range between $\approx 0.4$ and $\approx
0.3$, as the spin varies from $s=1/2$ to $s=3$.  Finally, it is also
unclear how to estimate the uncertainties involved in these procedures
and thus how to interpret the spread of exponent estimates, which
might be related to a strong spin-dependence of the slowly decaying
corrections appearing in eq.(\ref{glog}). Other prescriptions to study
the exponent of the logarithmic corrections, do not lead to better
results.

\subsection{ The critical exponents of the higher susceptibilities}

\begin{figure}[tbp]
\begin{center}
\leavevmode
\includegraphics[width=3.37 in]{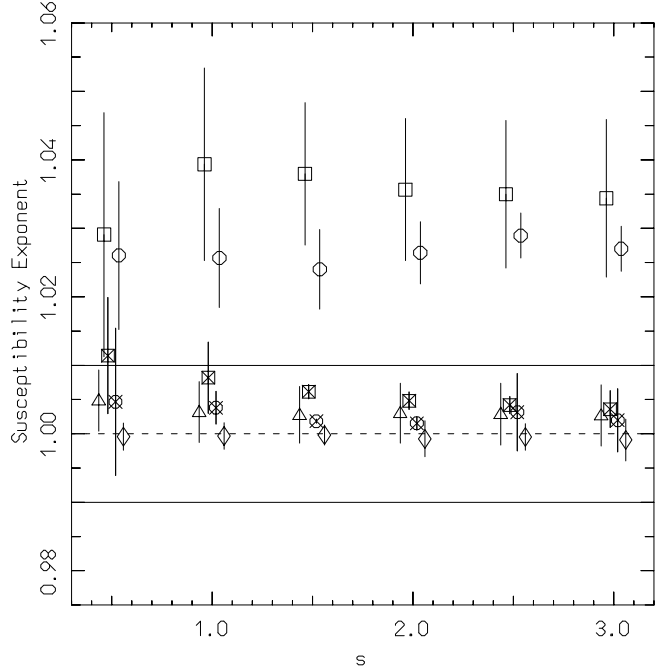}
\caption{ \label{gamma456} The exponent $\gamma$ of the susceptibility
vs the value $s$ of the spin, in the case of the spin-$s$ Ising models
on the h4sc, h4bcc, h5bcc and h6bcc lattices. For each value of the
spin, the different estimates and their uncertainties are made more
visible by slightly shifting their abscissas to avoid superpositions
of the symbols. The exponents are denoted by open squares in the case
of the h4sc lattice, or by open circles in the case of the h4bcc
lattice. If the logarithmic correction to the leading critical
singularity expected in 4D is canceled out from the susceptibility
expansion, we are led to the estimates represented by crossed open
squares in the case of the h4sc lattice and by crossed open circles in
the case of the h4bcc lattice. The estimates on the h5bcc and the
h6bcc lattices are represented by open triangles and open rhombs
respectively. The dashed horizontal line represents the expected value
$\gamma_{MF}=1$ of the exponent, and the continuous lines indicate a
$1\%$ relative deviation from that value.}
\end{center}
\end{figure}

For each model under study, we have computed the exponent $\gamma$ of
the susceptibility by second- or third-order DAs biased with our
estimate of the inverse critical temperature, namely by resorting to
the standard prescription of imposing that  the approximants  are singular 
at the values of $K_c$ reported in our tables
\ref{tabkc} and \ref{tabkcfi4} and then computing the exponents.  For
$d>4$, it is rigorously proved\cite{aize,fro} 
that the systems must exhibit a MF
critical behavior (with non trivial subleading corrections). Let us
 discuss first how our numerical tools perform in the 5D and 6D
cases.  We shall then argue that the differences between the features of this
computation and those of the 4D case can be simply ascribed to the
expected presence of a multiplicative logarithmic correction to the
dominant MF power behavior.  In Fig.\ref{gamma456}, we have plotted
our estimates of the exponent of the ordinary susceptibility vs the
spin in the case of various spin-$s$ Ising models for $d=4,5,6$. For
$d>d_c$, our estimates reproduce to a very good accuracy the expected
value $\gamma_{MF}=1$, so that the small deviations from this value
can be safely viewed as only 
 the residual effects of the confluent corrections to
scaling. These deviations also show the expected decreasing size as
the dimensionality of the system increases. Moreover the critical
universality, i.e. the independence of the exponents on the
interaction structure, is well verified.  On the contrary, at the
upper critical dimension our calculations yield ``effective''
exponents larger than unity by $\approx 3 \%$.  We can interpret this
result as an indication that the leading critical singularity of the
susceptibility is slightly stronger than a pure MF singularity so that it
might indeed contain the logarithmic factor predicted by the RG, which
is detected by the DAs as a power-like factor with a very small
exponent.  This is confirmed by observing that, if the expected
logarithmic singularity is canceled by dividing out from the
susceptibility the $ {\rm ln}(1-K/K_c)^{1/3}$ correction factor, the
resulting estimate of the exponent $\gamma$ generally gets within
$\approx 0.5 \%$ of the MF value. Thus the deviations are reduced to a
smaller size and become compatible with the effects of the corrections 
 to scaling.

 Very accurate  estimates can be obtained also for the  differences $D_n$
 between the exponents of $\chi_{2n}(K)$ and  $\chi_{2n-2}(K)$  
\begin{equation}
D_n= \gamma_{2n}- \gamma_{2n-2}= 2 \Delta_{MF}=3
\label{dn}
\end{equation}
They can be computed from the ratios $\chi_{2n}(K)/\chi_{2n-2}(K)$ by
second- or third-order DAs biased with the critical inverse
temperature. In the 4D case, we should not expect any effects from the
logarithmic factors appearing in the leading singular behavior
eq.(\ref{2ncorras4}) of the higher susceptibilities, since these
factors cancel in the above indicated ratios. Instead of the results
of the biased prescription, we prefer to show here the estimates from
a computation by the unbiased ``critical point renormalization''
method\cite{hunterbaker}.  This procedure consists in determining the
difference $D_n$ of eq.(\ref{dn}) from the exponent of the singularity
in $x=1$ of the series $\sum a_r x^r$ with coefficients
$a_r=c_r^{2n}/c_r^{2n-2}$, where $c_r^s$ is the $r$-th coefficient of
the expansion of $\chi_{s}(K)$.  The biased DA calculation of the
$D_n$, mentioned above, gives quite comparable results, so that it is
not necessary to report the corresponding figures.

\begin{figure}[tbp]
\begin{center}
\leavevmode
\includegraphics[width=3.37 in]{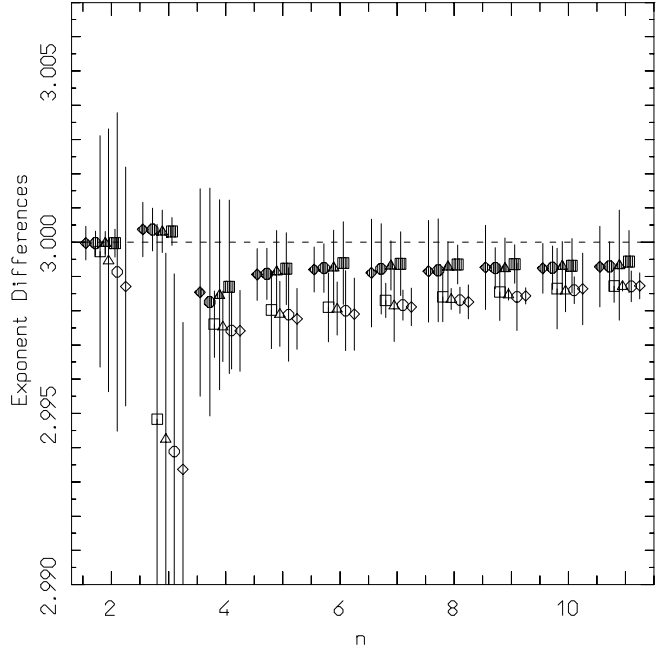}
\caption{ \label{diffga_fi4h56bcc} The unbiased estimates of the
 exponent differences $D_n=\gamma_{2n}-\gamma_{2n-2}$ for the
 susceptibilities $\chi_{2n}$ and $\chi_{2n-2}$, for $n=2,3,...11$
 plotted vs $n$, in the case of the scalar-field model on the h5bcc
 and h6bcc lattices.  In the case of the h5bcc lattice, for each value
 of $n$ we have reported a cluster of the four estimates of $D_n$
 corresponding to the values $g=0.9$ (open squares), $g=1.1$ (open
 triangles), $g=1.3$ (open circles), $g=1.5$ (open rhombs) of the
 self-coupling of the field. The same quantities for the h6bcc lattice
 are represented by the corresponding black symbols.  Although they
 correspond to the same value of $n$, the symbols within each cluster
 are slightly shifted apart to avoid cluttering and keep the spread of
 each estimate visible.  The dashed horizontal line represents the
 expected value $2\Delta_{MF}=3$ of twice the gap exponent.  }
\end{center}
\end{figure}

\begin{figure}[tbp]
\begin{center}
\leavevmode
\includegraphics[width=3.37 in]{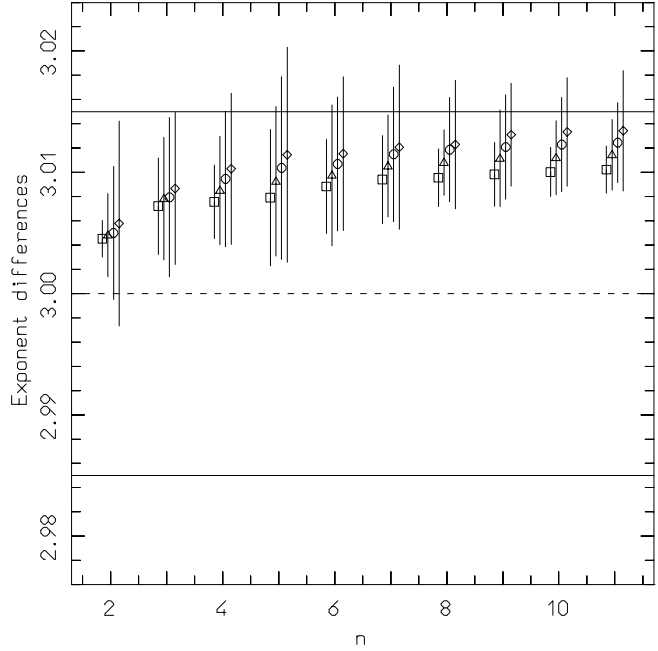}
\caption{ \label{diffga_fi4h4bcc} The unbiased estimates of the
exponent differences $D_n=\gamma_{2n}-\gamma_{2n-2}$ of the
susceptibilities $\chi_{2n}$ and $\chi_{2n-2}$, for $n=2,3,...11$
plotted vs $n$, in the case of the scalar-field model on the h4bcc
lattice.  For each value of $n$ we have reported a cluster of the four
estimates of $D_n$ corresponding to the values $g=0.9$ (squares),
$g=1.1$ (triangles), $g=1.3$ (circles), $g=1.5$ (rhombs) of the
self-coupling of the field.  Although they correspond to the same
value of $n$ the symbols within each cluster are slightly shifted
apart to avoid cluttering and keep the spread of each estimate
visible.  The dashed horizontal line represents the expected value
$2\Delta_{MF}=3$ of twice the gap exponent.  The continuous horizontal
lines indicate a relative deviation of $0.5 \%$ from the expected
value.}
\end{center}
\end{figure}

\begin{figure}[tbp]
\begin{center}
\leavevmode
\includegraphics[width=3.37 in]{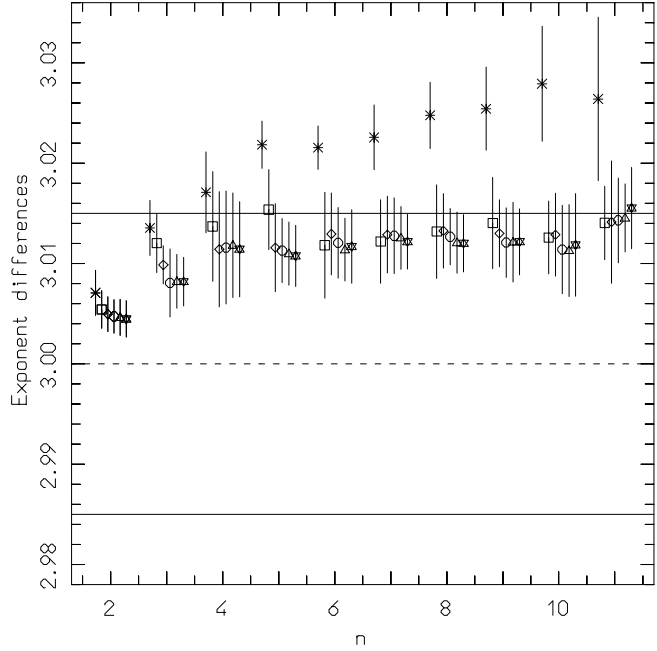}
\caption{ \label{diffga_Isi_h4bcc} The unbiased estimates of the exponent
  differences $D_n$ plotted vs $n$ in the case of the Ising model on
  the h4bcc lattice for the following values of the spin: $s=1/2$
  (asterisks) $s=1$ (open squares), $s=3/2$ (open rhombs), $s=2$ (open
  circles), $s=5/2$ (open triangles), $s=3$ (open stars). The
  horizontal dashed line and the continuous lines have the same meaning
  as in Fig.\ref{diffga_fi4h4bcc}. }
\end{center}
\end{figure}

\begin{figure}[tbp]
\begin{center}
\leavevmode
\includegraphics[width=3.37 in]{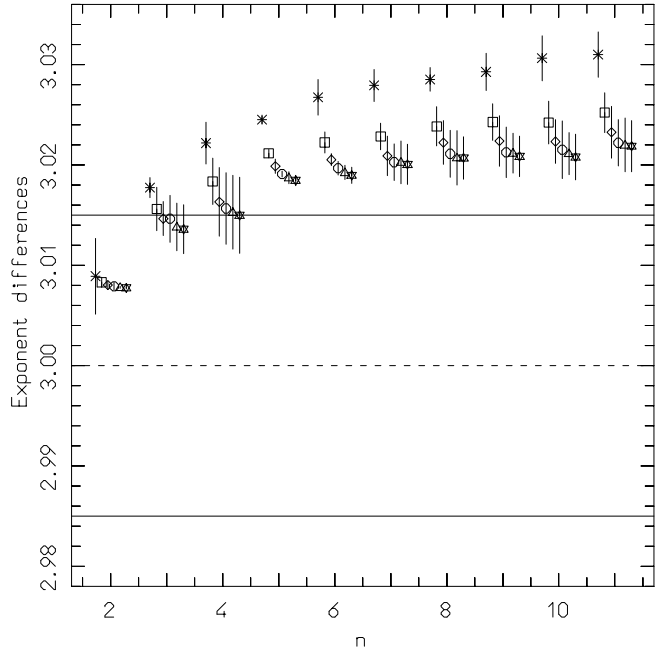}
\caption{ \label{diffga_Isi_h4sc} Same as Fig.\ref{diffga_Isi_h4bcc} 
 but for the Ising model on the h4sc lattice.
  The estimates of the exponent differences 
$D_n$  are plotted vs $n$ for spin  $s=1/2$ (asterisks)
 $s=1$ (open squares), $s=3/2$ (open rhombs), $s=2$ (open circles), 
$s=5/2$ (open triangles), $s=3$ (open stars).  The
  horizontal dashed line and the continuous lines have the same meaning
  as in Fig.\ref{diffga_fi4h4bcc}  }
\end{center}
\end{figure}

  The quantities $D_n$ with $n=2,3,...11$, obtained by the unbiased
method in the case of the the scalar-field model on the h5bcc and
h6bcc lattices, for several values of the coupling $g$, are plotted vs
$n$ in Fig.\ref{diffga_fi4h56bcc}.  The same computations for the
spin-$s$ Ising models with various values of the spin on the h5bcc and
h6bcc lattices yield completely similar results and therefore we do
not report the corresponding figure.  Our estimates of $D_n$ agree,
generally within $0.1 \%$, with the expected value
$2\Delta_{MF}=3$. Thus the small size of these deviations from the MF
value suggests that they can safely be related with the confluent
corrections to scaling.  The critical universality is also well
verified.  On the other hand, our results in the 4D case reported in
Fig.\ref{diffga_fi4h4bcc} in the case of the scalar-field model on the
h4bcc lattice, those reported in Fig.\ref{diffga_Isi_h4bcc} for the
Ising model on the h4bcc lattice and those of
Fig.\ref{diffga_Isi_h4sc} for the same system in the case of the h4sc
lattice, show relative deviations from the value of $2\Delta_{MF}$,
five times larger than those in $d>4$ dimensions (i.e. of the order of
$0.5 \%$), but still sufficiently small to reflect only the residual
influence of the expected subleading logarithmic corrections to the
critical behavior.

\subsection{ Universal combinations of critical amplitudes} 

\begin{figure}[tbp]
\begin{center}
\leavevmode
\includegraphics[width=3.37 in]{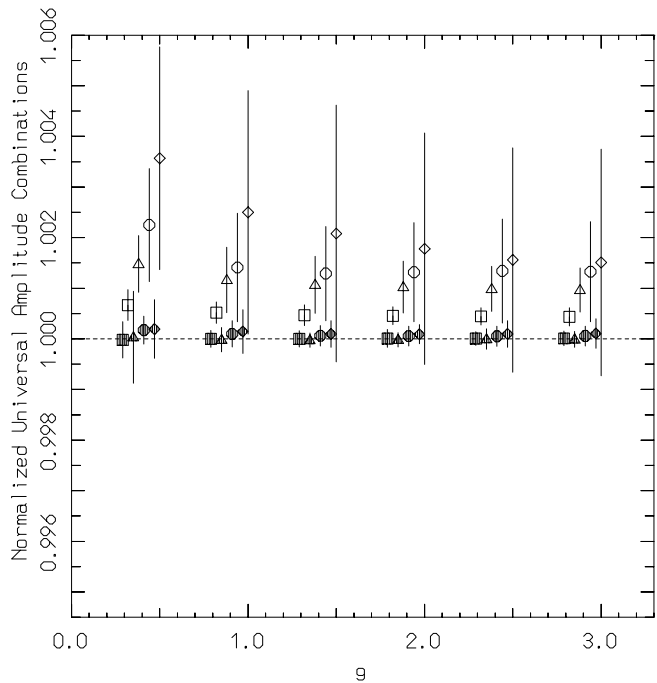}
\caption{ \label{IsuImfh56bcc} The ratios ${\cal Q}_{2r+4}={\cal
I}^+_{2r+4}/\hat {\cal I}^+_{2r+4}$ with $r=1,2,3,4$ vs the value $s$
of the spin for the Ising model on the h5bcc and h6bcc lattices.
For each value of $s$ the various
symbols are slightly shifted apart to avoid superpositions and to keep
the spread of each estimate visible.  
In the case of the h5bcc lattice we have represented ${\cal Q}_{6}$ by 
open squares, ${\cal Q}_{8}$ by  open
triangles, ${\cal Q}_{10}$ by open circles, ${\cal Q}_{12}$ by open rhombs.  
The same ratios for the h6bcc lattice are represented
by the corresponding black symbols.  The
  horizontal dashed line represents the expected value of the ratios.  }
\end{center}
\end{figure}

\begin{figure}[tbp]
\begin{center}
\leavevmode
\includegraphics[width=3.37 in]{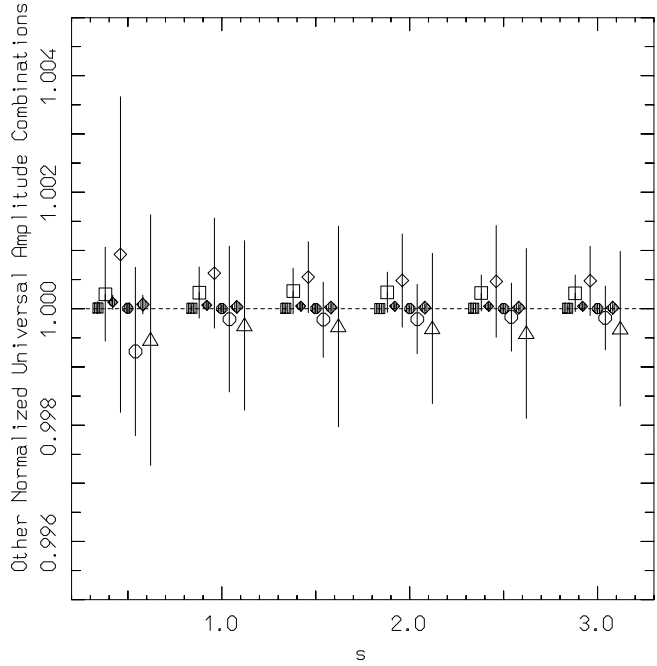}
\caption{ \label{nuoveuccah56bcc} The  ratios $ {\cal
R}^+_{8}$ (open squares),  $ {\cal
R}^+_{10}$ (open circles), 
$ {\cal S}^+_{10}$ (open rhombs),and $ {\cal S}^+_{12}$ (open triangles) vs
the spin $s$ in the case of the Ising model with various values of the
spin on the h5bcc lattice. The corresponding black symbols represent
the estimates of the same quantities on the h6bcc lattice. For each
value of $s$ the various symbols are slightly shifted apart to avoid
superpositions and to keep the spread of each estimate visible. The
  horizontal dashed line represents the expected value of the ratios. }
\end{center}
\end{figure}

For $d>4$, using second- or third-order DAs, the first few terms of
 the sequence of the UCCAs ${\cal I}^+_{2r+4}$, ${\cal A}^+_{2r+4}$
 and ${\cal B}^+_{2r+8}$ can be evaluated to a good accuracy, by
 extrapolating to $K=K_c^-$ the estimates of the functions ${\cal
 I}_{2r+4}(K)$, ${\cal A}_{2r+4}(K)$ and ${\cal B}_{2r+8}(K)$. For
 convenience, we have introduced the ratios of these quantities to
 their MF values, and denoted them by ${\cal Q}_{2r+4}$, ${\cal
 R}_{2r+4}$ and ${\cal S}_{2r+8}$, respectively.  In
 Fig.\ref{IsuImfh56bcc}, we have reported our estimates for the ratios
 ${\cal Q}_{6}$, ${\cal Q}_{8}$, ${\cal Q}_{10}$ and ${\cal Q}_{12}$
 vs the value $s$ of the spin for Ising models on the h5bcc and h6bcc
 lattices.  In complete agreement with the proven MF nature of the
 critical behavior, these ratios generally equal unity, within the
 accuracy expected from our approximations that, in this case, allows
 not only for the influence of the confluent corrections to scaling,
 but also for the uncertainties in the estimates of the critical
 temperatures needed to bias the calculations.  Correspondingly, the
 critical limits of the ratios $r_{2n}(K)$ vanish and the equation of
 state takes the MF form.  Quite similar results are shown in
 Fig.\ref{nuoveuccah56bcc} for the other normalized UCCAs ${\cal
 R}^+_{8}$, ${\cal R}^+_{10}$, ${\cal S}^+_{10}$ and ${\cal S}^+_{12}$
 which are plotted vs the spin $s$ for Ising models with various
 values of the spin in the case of the h5bcc and h6bcc lattices.

\begin{figure}[tbp]
\begin{center}
\leavevmode
\includegraphics[width=3.37 in]{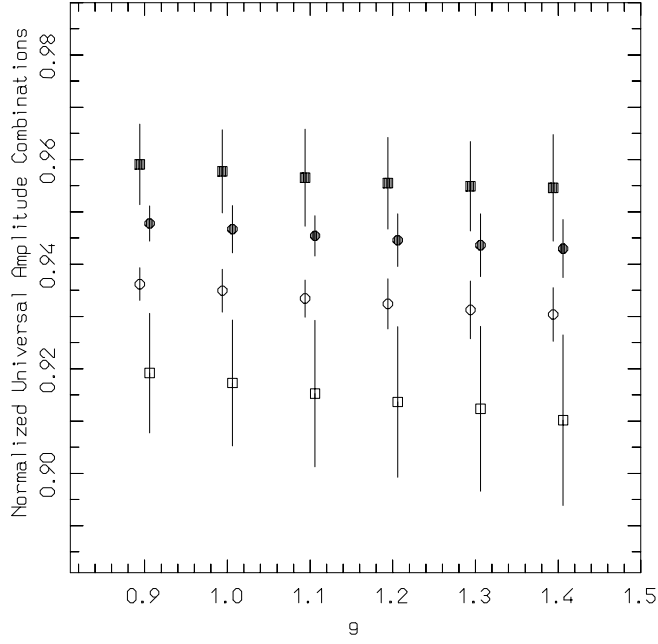}
\caption{ \label{I6I8suImf_fi4_h4sc_h4bcc} The ratios ${\cal
Q}_6={\cal I}^+_6/\hat {\cal I}^+_6$ (black squares) and ${\cal
Q}_8={\cal I}^+_8/\hat {\cal I}^+_8$ (open squares) for the scalar
field model on the h4sc lattice vs the coupling constant $g$ of the
field. The same quantities ${\cal Q}_6$ (black circles), and ${\cal
Q}_8$ (open circles) vs the coupling constant $g$ for the scalar-field
model on the h4bcc lattice. Like in the preceding figure, symbols
associated to the same value of $g$ are slightly shifted in order to
avoid superpositions of the error bars. }
\end{center}
\end{figure}

\begin{figure}[tbp]
\begin{center}
\leavevmode
\includegraphics[width=3.37 in]{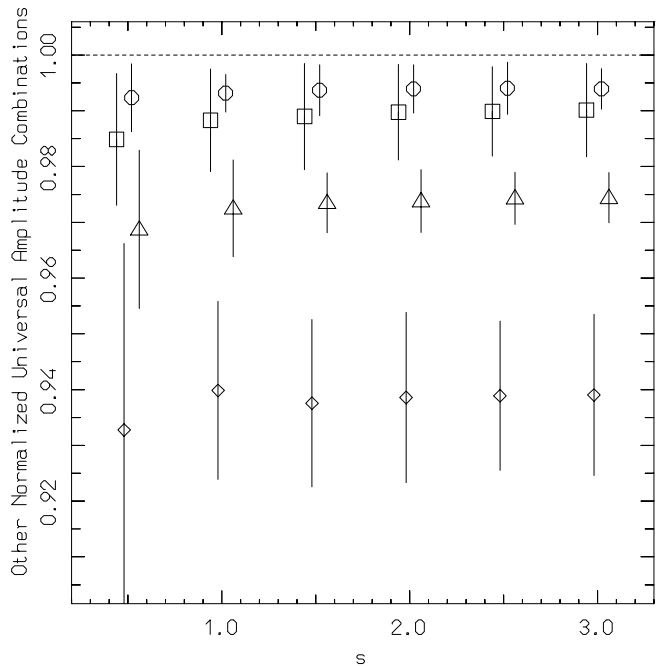}
\caption{ \label{nuoveuccah4bcc} The ratios $ {\cal R}^+_{8}$ (open
 squares), $ {\cal R}^+_{10}$ (open circles), $ {\cal S}^+_{10}$ (open
 rhombs) and $ {\cal S}^+_{12}$ (open triangles) plotted vs the spin
 $s$ in the case of the Ising model on the h4bcc lattice.Like in the
 preceding figure, symbols associated to the same value of the spin
 are slightly shifted in order to avoid superpositions of the error
 bars.  }
\end{center}
\end{figure}

Also in the 4D case, as shown in Fig.\ref{I6I8suImf_fi4_h4sc_h4bcc}
 for the first few UCCAs defined by eq.(\ref{watsoncr}) in the case of
 the scalar-field model, and in Fig.\ref{nuoveuccah4bcc} for a few
 UCCAs defined by eqs.(\ref{watsonar}) and (\ref{watsonbr}) in the
 case of the spin-$s$ Ising model, the various quantities probably
 have been evaluated with reasonable accuracy, because the logarithmic
 factors, expected to appear in the leading critical behavior of the
 higher susceptibilities, cancel in the ratios defining the UCCAs. As
 a consequence, the uncertainties in the critical temperatures and the
 influence of the corrections to scaling should still be considered as 
 the main sources of error.  However, we observe that generally the
 first few ratios ${\cal Q}_{2r+4}$, ${\cal R}_{2r+4}$ and ${\cal
 S}_{2r+8}$ are slightly, but definitely smaller than unity.  We can
 imagine two possible explanations of this result: either the
 deviations from unity have to be related only to (unlikely) residual
 effects of the logarithms in the leading and subleading behavior of
 the higher susceptibilities, or the UCCAs are accurately estimated
 and they really do not take their MF values. Whatever the case, it is
 clear that also these data on the UCCAs confirm that, consistently
 with the RG predictions, the critical behavior in 4D is not MF-like.

\section{Conclusions}
By analyzing our HT expansions of the zero-field
 higher-susceptibilities, extended through order 24, in the case of
 the $N=1$ lattice scalar-field models and of the spin-$s$ Ising
 systems, we have provided further numerical evidence consistent with
 the critical behavior predicted by the RG in this class of models.

  We have estimated the critical exponents of the ordinary and the
 higher susceptibilities and the values of a class of universal combinations of
 their critical amplitudes, which determine the form of the critical
 equation of state and are presently inaccessible by other
 computational methods. In 4D, the results of our analysis suggest
 that, within a good approximation, the critical exponents and this
 class of UCCAs, show small, but definitely nonvanishing deviations
 from their values in the MF approximation.  For the UCCAs, this fact
 had been already predicted long ago also within the RG formalism, by
 showing\cite{brezin} that, at the upper critical dimension, at least
 one of the quantities in the above mentioned class does not take the
 MF value. More generally, in 4D the deviations from the MF critical
 behavior are compatible with the small effects associated to the
 logarithmic corrections predicted by the RG.  Our direct numerical
 checks concerning in particular the exponents of the logarithmic
 corrections to the dominant power behavior of the higher
 susceptibilities have only a rather limited accuracy, due to the
 modest sensitivity of the DAs to the logarithmic singularities,
 either in the leading behaviors and in the confluent corrections.

Quite on the contrary, the same kind of analysis performed on five-
and six-dimensional lattices, shows no numerical evidence of
deviations from the leading classical behavior by an extent larger
than the expected numerical uncertainties: both the exponents and the
UCCAs appear to take the MF values within a high approximation, so
that the RG predictions concerning the triviality property are rather
convincingly confirmed.

\section{Acknowledgements}
We thank Ian Campbell, Michael E. Fisher, Ralph Kenna and Ulli Wolff
 for their patience in reading and commenting a preliminary draft of this
 paper. The hospitality and support of the Physics Depts. of
 Milano-Bicocca University and of Milano University are gratefully
 acknowledged.  Partial support by the MIUR is also acknowledged.

\end{document}